# DEFUZZIFICATION METHOD FOR A FASTER AND MORE ACCURATE CONTROL

S. Sanyal T.I.F.R. Homi Bhaba Road, Bombay -5, India,
S. Iyengar, A.A. Roy, N.N. Karnik, N.M. Mengale, S.B. Menon,
Victoria Jubilee Technical Institute, Matunga, Bombay-19, India,
Wu Geng Feng, Dept of Electrical & Electronics Enng.,
USTC, Hefei 230026, Anhui, PRC.

## INTRODUCTION

Today manufacturers are using fuzzy logic in everything from cameras to industrial process control. Fuzzy logic controllers are easier to design and so are cheaper to produce. Fuzzy logic captures the impreciseness inherent in most input data. Electromechanical controllers respond better to imprecise input if their behavior was modeled on spontaneous human reasoning.

In a conventional PID controller, what is modeled is the system or process being controlled, whereas in the Fuzzy logic controller, the focus is the human operator behavior. In the first case, the system is modeled analytically by a set of differential equations and their solutions tells the PID controllers how to adjust the system's control parameters for each type of behavior required [3]. In the Fuzzy controller these adjustments are handled by a Fuzzy rule based expert system. A logical model of the thinking process a person might go through in the course of manipulating the system. This shift in focus from process to person involved changes the entire approach to automatic control problems.

## ABSTRACT

Fuzzy logic is achieved by formulating a rule base which is based on experience gathered by human operators. Those systems which cannot be modeled mathematically benefit most from Fuzzy control strategy since the imprecise data can be captured by using linguistic data in the rule-base.

Fuzzy logic has certain disadvantages. The number of computations required for arriving at a certain output for a given input is very large and thus the system response is sluggish. Therefore to adapt Fuzzy systems to real time applications we need to use faster algorithms and/or parallel processing. Also the process of defuzzification required to produce single output may lead to errors which can undo the advantages of fine control normally achievable by Fuzzy logic. Special analytical techniques ensure that the complications of the defuzzification process are simplified; the method used retains the desired features of fuzzy control.

This paper aims to present a comparison between the conventional fuzzy controller and a fuzzy logic controller based on the techniques mentioned above.

## EXPERIMENTAL

1. The problem sought to be tackled is that of an inverted pendulum. In this classical problem a pole is attached to a vehicle by a hinge such that from an upright position it can fall only to right or left. The role is to monitor the pole's angular position and speed and move the vehicle left or right accordingly, to keep the pole upright.
The conventional fuzzy control approach for triangular distribution of primary sets and the suggested approach using parabolic distribution have been implemented in software. The observations from the programs are discussed in this paper.

2. Rule Base

The rule base sufficient to balance the pendulum is shown in table 1 [3]. The rules can be written based on logical reasoning to prevent the pendulum from falling. The exact set of rules can be tailored to suit the dynamics of the physical components, required robustness and range of operating conditions.
The rule-base of the two programs is kept identical so that the difference in control action can be attributed to the suggested approach.

|    | NB | NM | NS | ZE | PS | PM | PB |
|----|----|----|----|----|----|----|----|
| NB |    |    |    |    |    |    |    |
| NM |    |    |    |    |    |    |    |
| NS |    |    | NS |    | ZE |    |    |
| ZE |    | NM |    | ZE |    | PM |    |
| PS |    |    | ZE |    | PS |    |    |
| PM |    |    |    |    |    |    |    |
| PB |    |    |    |    |    |    |    |

Table 1. Rule-base

## CONVENTIONAL FUZZY CONTROL

A Fuzzy logic based control system involves the following steps – fuzzification of the input variables, firing of rules in the rule-base and finally defuzzification to yield the control action. Conventional fuzzy systems employ triangular curves for fuzzification, primarily because of lesser computations involved. A typical set of membership functions spanning the universe of discourse is chosen (Fig 1). The linguistic variables are represented by isosceles triangles. Triangular membership functions are easily represented using straight line equations.
The equations defining one membership curve are:

$y = (x-a)/d$ and
$y = 1-(x-a)/d$

where $a$ = the abscissa of the peak of the membership curve and $d$ = the constant difference between peaks of the two consecutive curves.

Seven such primary sets are used viz., NB, NM, NS, ZE, PS, PM, PB and they span a normalized range from -1 to +1. The fuzzified input data fire all rules in the rule-base. Since there are two input variables (angle and angular velocity), the rules employ union and intersection of the fuzzy sets. Based on the value returned by each rule, the corresponding membership curve in the output control set is clipped. This method is known as MAX-MIN method transforms these curves into trapezoids (Fig 2).

Most control systems use the centroid defuzzification method. In this defuzzification method, the area of each consequent set is multiplied by the domain values passing through its centre. The sum of these products is then divided by the sum of the area values of all the sets. This yields the normalized defuzzification values which can be mapped to yield the control force.

PARABOLIC FUZZY CONTROL

As the name implies, primary sets are represented by parabolic functions in this controller program (Fig 3).

1. Degrees of fuzziness and their significance.

Fuzzy logic differs from conventional logic in that it does not tend to avoid the negative; rather, it encourages the cohabitation of the positive and the negative. This means that the Fuzzy membership functions must be such that the areas of intersection with their respective complimentary functions must be as small as possible. The intersection-area between a membership function and its compliment is represented by a numerical value that after due normalization is called the degree of fuzziness of that membership function. A good membership function has a large degree of fuzziness and a small area of intersection with its compliment function.

Selecting a 'good' membership function for representing a linguistic variable in a rule in a rule-base builds in robustness into the fuzzy control system. This is because the complimentary function becomes a part of the neighbouring linguistic variable's curve. This means that for a given input value, a membership value of NS is very close to the membership value of its neighbouring curve NM in the region of overlap and so on and so forth. Thus the concepts of fuzzy logic are best captured by such 'good' functions and the resulting membership values are the most accurate representations of the fuzzification of the input value and this in turn means that a control action that results from the membership value will also be the most accurate. A fuzzy system that uses 'good' functions will also be robust because the fuzzified output curves will have multiple peaks.

2. Degree of fuzziness for parabolic membership functions[4].

The following representation shows that parabolic membership functions have more degree of fuzziness than equivalent triangular membership functions:

a. TYPE OF MEMBERSHIP FUNCTION
Triangular.
EQUATIONS
$y = (x-a)/D$          $a <= x <= b$,
$y = -(x-c)/D$          $b < x <= c$,
$D = b-a = c-b$,
AREA OF INTERSECT WITH COMPLIMENT
$D/2$
DEGREE OF FUZZINESS
0.25(Intermediate)

b. TYPE OF MEMBERSHIP FUNCTION
Parabolic - I.
EQUATIONS
$y = 2(x-a)^2/D^2$          $a <= x <= (a+b)/2$,
$y = 1-2(x-b)^2/D^2$          $(a+b)/2 < x <= (b+c)/2$,
$y = 2(x-c)^2/D^2$          $(b+c)/2 < x <= c$
$D = b-a = c-b$,

AREA OF INTERSECT WITH COMPLIMENT
$2D/3$
DEGREE OF FUZZINESS
0.16(Least)

c. TYPE OF MEMBERSHIP FUNCTION
Triangular+Parabolic( Mixed)
EQUATIONS
$y = (x-a)/D$          $a <= x <= (a+b)/2$,
$y = \frac{1}{2}+2(x-(a+b)/2)^2/D^2$          $(a+b)/2 < x <= b$,
$y = \frac{1}{2}+2(x-(b+c)/2)^2/D^2$          $b < x <= (b+c)/2$,
$y = -(x-c)/2$          $(b+c)/2 < x <= c$
$D = b-a = c-b$,

AREA OF INTERSECT WITH COMPLIMENT
$5D/12$
DEGREE OF FUZZINESS
0.29(Intermediate)

TYPE OF MEMBERSHIP FUNCTION
Parabolic - II.
EQUATIONS
$y = \frac{1}{2}-2(x-(a+b)/2)^2/D^2$          $a <= x <= (a+b)/2$,
$y = \frac{1}{2}+2(x-(a+b)/2)^2/D^2$          $(a+b)/2 < x <= b$,
$y = \frac{1}{2}+2(x-(b+c)/2)^2/D^2$          $b < x <= (b+c)/2$,
$y = \frac{1}{2}+2(x-(b+c)/2)^2/D^2$          $(b+c) < x <= c$,
$D = b-a = c-b$,

AREA OF INTERSECT WITH COMPLIMENT
$D/3$
DEGREE OF FUZZINESS
0.33(Highest)

*a and c are range limits for the curves though the fuzzy sets themselves have a range of [-1,1], b is the abscissa of the peak of the curve.*

DEGREE OF FUZZINESS (f) is calculated from AREA OF INTERSECTION WITH COMPLIMENT ($A_x$) as:
$f = 1 - (A_x/(c-a))$

Seven primary types of Parabolic II type are chosen. viz., NB, NM, NS, ZE, PS, PM, PB and they span a normalized range from -1 to +1.

Following the process fuzzification rule base is fired yielding the membership values in the consequent sets. The process of obtaining the consequent sets involves the process of scaling down the primary sets so that the height of each parabola is now equal to the membership value of the input variable in each primary set. After scaling, the curves retain the properties of the parabolas (Fig 4). Scaling increases the accuracy of the process of area calculation. This is because the approximations involved in the calculations of the output control actions in the program using triangular primary sets are now avoided. This leads to increased accuracy in calculations of the final defuzzified output value which in turn gives a more accurate and closer control. The calculation of the area of the individual parabolas can be done faster due to reduction in computational overheads. The area of each parabola is calculated in terms of the scaling factor for each curve. A lot of calculations are done a priori which results in faster calculation of the area and finally the output control action.

The output action calculated is now in normalized form which is mapped into real world values and a feedback loop is used to simulate the response of the physical system. The resultant angle is fed back from a procedure that simulates the operation of an inverted-pendulum system [2] and this resultant angle is then used to calculate a new force using the program. The loop is repeated till the stability is achieved.

ACKNOWLEDGEMENTS

We are extremely grateful to Prof P.V.S. Rao, Head, Computer Systems and Communications Group, T.I.F.R.,

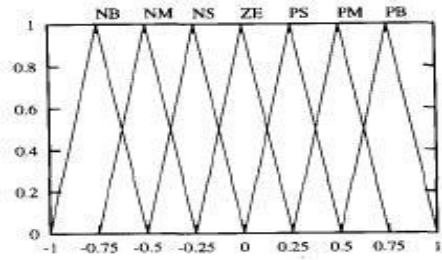
Fig 1. Primary Sets For Conventional Fuzzy Control

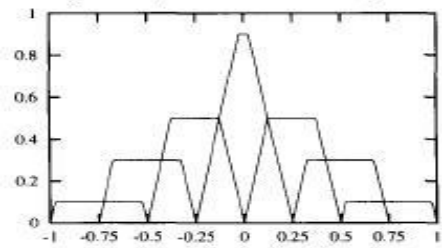
Fig. 2 Consequent Sets For Conventional Fuzzy Control

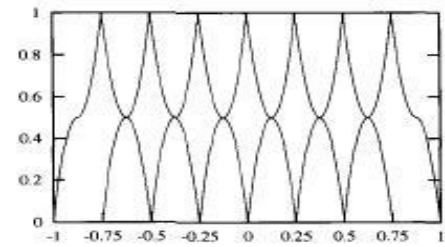
Fig 3. Primary Sets For Parabolic Fuzzy Control

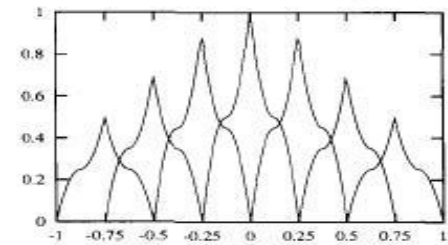
Fig 4. Consequent Sets For Parabolic Fuzzy Control